\documentclass[11pt,twoside]{article}
\usepackage{./asp2014}

\resetcounters

\begin{document}

\title{Science with an ngVLA: Stellar Activity on Red Giant and Supergiant Stars: Mass Loss and the Evolution of
the Stellar Dynamo.}

\author{Graham M. Harper}
\affil{Center for Astrophysics and Space Astronomy, University of Colorado Boulder, CO, USA; \email{graham.harper@colorado.edu}}

\paperauthor{Graham M. Harper}{graham.harper@colorado.edu}{0000-0002-7042-454180}{University of Colorado}{Center for Astrophysics and Space Astronomy}{Boulder}{CO}{}{USA}

\begin{abstract}
In this Chapter we examine the role of the ngVLA to further our understanding of the different manifestations of convective or turbulence-driven stellar activity on red giant and supergiant stars.  
The combination of high spatial resolution and high sensitivity will enable the ngVLA to significantly improve our understanding of the processes that dissipate energy in the extended atmospheres of cool evolved stars, and drive ubiquitous stellar outflows.  The high spatial resolution will enable us to image the surfaces of nearby red supergiants, and to measure the atmospheric extent of red giants. Multi-frequency observations will permit thermal continuum tomography on the largest angular diameter stars, providing key empirical data to test theoretical models. The complementary frequencies and similar spatial resolutions of the ngVLA and ALMA will be a powerful synergy.
\end{abstract}

\section{Context}

Cool stars have vigorous sub-photospheric convection zones that lead to turbulent motions 
in their photospheres, and in the layers above.  
The interaction of surface 
magnetic fields \citep[e.g.,][]{auriere2015} and dynamic ionized plasma leads to complex 
phenomena including starspots, active regions, chromospheres, and stellar outflows. These manifestations of 
{\it Stellar Activity} \citep{schrijver2000} are not well understood but have important 
astrophysical consequences, ranging from the effects of mass loss
on galactic structure and chemical evolution, 
the nature of space-weather for exo-planets orbiting red giants, to the effects of cool supergiant 
mass-loss history on
the interpretation of early-time 
supernovae spectra \citep[e.g.,][]{dessart2017}. Theoretical models have not been able to reproduce observational 
signatures of stellar activity, and progress is being led by state-of-the-art observations 
made at the highest spatial resolution and with the greatest sensitivity.

Thermal radio continuum observations are powerful diagnostics in this context because the 
atomic cross-sections are very accurately known \citep{hummer1988} and the emission source term is the 
Planck function, 
which is linear in electron temperature.  For example, resolved VLA images provided a ground 
breaking discovery and new insights into red supergiant (RSG) outflows with the finding by 
\cite{lim1998} that Betelgeuse's extended atmosphere is cooler and much less ionized 
than predicted by leading theoretical and semi-empirical models of the time \citep[see][]{ha1984, 
harper2001}. 

However, red giants and supergiants are not strong radio emitters and have relatively small angular sizes
so that future progress requires greater sensitivity {\it and} higher spatial resolution. 
The ngVLA will be able to provide both of these. 

\section{Mass Loss}

Red giants and RSGs return mass that is enriched by nuclear processing
back into the interstellar medium, supplying giant molecular clouds that may host the next
generation of star formation, and leading to new stars and their exo-planets. The rate at which mass is returned by cool stars
is, to a first approximation, proportional to a star's surface area and the inverse cube of the
surface escape speed \citep{hfl83}. So low surface-gravity cool evolved stars with radii
$10-10^3\> {\rm R}_\odot$ currently inject more processed material than their main-sequence 
progenitors. However, the mechanisms that drive mass-loss for K through mid-M giants and
the yellow and RSGs are very poorly understood: They have too little dust or molecules in their
outflows for radiation driven winds; they have too little photospheric variability or pulsation to
have acoustic and shock driven winds \citep[e.g.,][]{arroyo2015}, and they are too cool for a Parker-like
thermal winds. Some form of magnetic activity, or a combination of
processes, is likely to be the cause of these ubiquitous mass outflows \citep{lamers1999}. Clearly 
a quantitative physical model for mass loss is lacking for these stars, and even for the 
later spectral-type pulsating asymptotic giant branch (AGB) 
stars the role of magnetic fields in driving and shaping the outflows is not understood.

\subsection{What Empirical Data do we Need to Make Progress?}

Outflows from cool evolved stars have terminal wind speeds, ${\mathsf v}_\infty$, that are typically a small fraction of the 
surface escape speed, ${\mathsf v}_{esc}$. This means that
{\bf most} of the energy that goes into driving the outflows, proportional to ${\mathsf v}_{esc}^2 + {\mathsf v}_\infty^2$, goes into overcoming the
star's gravitational potential, i.e., $\propto {\mathsf v}_{esc}^2$. Therefore, {\em the optimum region to study mass 
loss mechanisms is within the first few stellar radii where most of the energy goes into the wind and the thermodynamic signatures of the outflow 
driving mechanisms will be 
most apparent}.

Nearby red giants have photospheric angular
diameters, $\theta_\ast$, of $10-20$\,mas, and the two nearest RSGs (Betelgeuse: M2~Iab, and
Antares: M1.5~Iab) have $\theta_\ast\simeq 42-44$\,mas \citep{ohnaka2013,montarges2016}. The luminous Herschel's Garnet Star ($\mu$~Cep: M2~Ia) has $\theta_\ast=14.1$\,mas \citep{perrin2005},
the next nearest luminosity class Iab-Ib 
RSG (CE~Tau) has $\theta_\ast\simeq 10$\,mas \citep{cruzalebes2013}, and $\alpha^1$~Her (M5~Ib-II) has
$\theta_\ast=33$\,mas \citep{benson1993}.  Importantly, at 50\,GHz, the atmospheres of RSGs have 
angular extents about twice that of their photospheres, and the ngVLA, with baselines of 200\,km, would provide a resolution of
better than 10\,mas. This would match the highest ALMA spatial resolutions, for example, as
shown in Figure~1.

\begin{figure}[t]
\center
\includegraphics{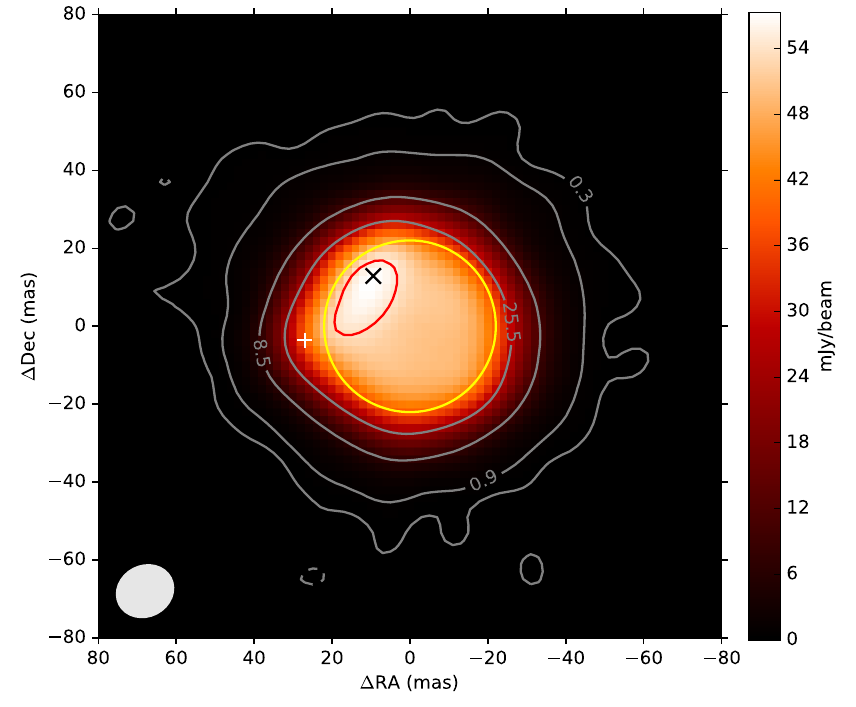}
\caption{338\,GHz ALMA map of Betelgeuse with a half-power beam width of $\simeq 15$\, mas (shown lower left),
from \cite{ogorman2017}. The yellow circle is the photospheric angular diameter measured with VLTI-PIONIER 
in H-band ($\simeq 1.6\>\mu$m) by \cite{montarges2016}. 
The larger angular extent seen at ALMA frequencies reveals the 
lower chromosphere and temperature minimum region. In contrast, the frequencies probed by the VLA sample the
more extended chromosphere and wind acceleration region. In this image localized excess emission
is marked with crosses. The ngVLA would be able to image the extended regions with the
same spatial resolution as ALMA, allowing us to build a tomographic map from the upper-photosphere 
out into the region where the outflow is initiated. Credit: O'Gorman et al., A\&A, Vol. 602, p. L11, 2017, 
reproduced with permission \copyright\,ESO.}
\end{figure}

This figure shows the 338\,GHz (0.89\,mm) thermal
continuum image of the RSG Betelgeuse obtained with ALMA with the longest (16\,km) baselines, providing 
a spatial resolution of 15\,mas \citep{ogorman2017}. 
At this wavelength the atmosphere of the star has an angular diameter of $\sim 55$\,mas, and this image shows 
the non-uniform
heating in the atmosphere that might be related to giant convection cells
or regions of enhanced magnetic fields.
The thermal cm-continuum opacity  $\kappa_\lambda \propto \lambda^{2.1}$, so observations at multiple wavelengths
map out the temperature and ion density distribution as a function of radius, providing vital clues to
the processes levitating and heating the stellar outflow. This requires that the ngVLA 200\,km
baselines support a range of observing frequencies.

\section{Dynamos in Cool Evolved Stars}

Most studies of solar-type  $\alpha\Omega$ dynamos have focused on main-sequence stars that
explore a relatively small parameter space of radius and mass, with a wide range of
rotation rates \citep[e.g.,][]{brun2017}. Red giants, however, provide a different challenge for dynamo theories because they
typically have low rotation rates but a much greater range of radii and convection zone depths. 
It is an open question as
to whether an $\alpha\Omega$ dynamo is maintained on the red giant branch or whether
a turbulent dynamo \cite[e.g.,][]{durney1993} is responsible for the generation of magnetic fields that
power the ultraviolet (UV) and X-ray emission, and drive stellar outflows.
Another important application for the long baselines of the ngVLA is to measure the extent of the
magnetically heated chromosphere of red giant stars. Spatial resolution breaks
the {\em spatial impasse} that limits what can be learned from disk integrated flux
diagnostics alone \citep[e.g.,][]{harper2013}.  There are some
recent clues that red giants show signs of magnetic cycles 
\citep{sennhauser2011} and non-uniform
optical brightness distributions \citep{richichi2018}, indications that magnetic phenomena might
be important in creating starspots and active regions.
High spatial resolution ngVLA observations would reveal the radio size and
uniformity of the chromosphere for the nearest red giant stars. This information
would help constrain the nature of stellar dynamos, and thus the shape of magnetic fields that
drive outflows from red giants. Such radio observations might also help determine
whether low-activity red giants, so called basal-flux stars, are heated by acoustic
shocks \citep{perez2014}. These would have compact chromospheres, but if
they too are heated by magnetic fields \citep{judge1998} they may have more extended atmospheres. There are
$12$ K and M non-Mira red giants with limb-darkened angular diameters $>$10\,mas obtained with the Mark III Stellar Optical Interferometer 
\citep{markiii} that would be
resolved at 50\,GHz with 200\,km baselines. These stars are given in Table~1 along with the MK spectral-types from
\cite{keenan1989}. The atmospheric electron density scale-heights are expected to be 
small compared to the stellar radii, so that the radio specific intensity distribution is, to first order, 
a top-hat and thus relatively easy to interpret. A sweep through radio continuum frequencies provides an opportunity
to build a tomographic map of the mean atmospheric temperature profile: ngVLA observations would sample the key 
chromosphere and wind regions, while synergistic ALMA observations would probe 
the temperature minimum and upper photosphere \citep[see Figures~1 and 2 of][]{ogorman2013}.

\begin{table}[!ht]
\caption{Stellar properties for red giants with $\theta_\ast>$10\,mas \citep{markiii}.}
\smallskip
\begin{center}
{\small
\begin{tabular}{rllcc}  
\tableline
\noalign{\smallskip}
HR  &  Name & Spectral-Type & Ang. Diam. (mas) & $T_{eff}$ (K) \\
\noalign{\smallskip}
\tableline
\noalign{\smallskip}
5340  & $\alpha$~Boo  & K1.5~III    & $21.37\pm 0.25$ & $4226\pm 53$ \\
5563  & $\beta$~Umi   & K4-~III     & $10.30\pm 0.10$ & $3849\pm 47$ \\
1457  & $\alpha$~Tau  & K5+~III     & $21.10\pm 0.21$ & $3871\pm 48$ \\
337   & $\beta$~And   & M0+~IIIa    & $13.75\pm 0.14$ & $3763\pm 46$ \\
6056  & $\delta$~Oph  & M0.5~III    & $10.47\pm 0.12$ & $3721\pm 47$ \\
911   & $\alpha$~Cet  & M1.5~IIIa   & $13.24\pm 0.26$ & $3578\pm 53$ \\
2216  & $\eta$~Gem    & M2~IIIa     & $11.79\pm 0.12$ & $3462\pm 43$ \\
8775  & $\beta$~Peg   & M2.5~II-III & $17.98\pm 0.18$ & $3448\pm 42$ \\
2286  & $\mu$~Gem     & M3~IIIab    & $15.12\pm 0.15$ & $3483\pm 43$ \\
4910  & $\delta$~Vir  & M3+~III     & $10.71\pm 0.11$ & $3602\pm 44$ \\
921   & $\rho$~Per    & M4+~IIIa    & $16.56\pm 0.17$ & $3281\pm 40$ \\
6146  & g~Her         & M6-~III     & $19.09\pm 0.19$ & $3008\pm 37$ \\
\noalign{\smallskip}
\tableline\
\end{tabular}
}
\end{center}

\end{table}

\section{Synergies with other Observatories}

Free-free radio continuum diagnostics  
{\it directly} relate to the thermodynamic state of the atmospheres, and spatially-resolved
multi-wavelength observations provide measures of the temperature and electron (ion) density scale-heights.
The radio therefore provides the physical context with which to interpret optical and IR images made a similar high 
spatial resolutions. The 10\,mas spatial resolution obtainable with the ngVLA complements the spectro-interferometry 
and imaging currently being made with the {\it Very Large Telescope}, 
\citep[e.g.,][]{kervella2016, montarges2017}.

Another ngVLA synergy is that with UV spectroscopy. UV emission line fluxes and radio optical
depths are proportional to the emission measure, i.e.,
$$
F_{UV} \propto \int n_e n_H \> dR\> dA \>\>\>\>\>\>\>\>\> \tau_{radio} \propto \int n_e^2\> dR
$$   
where $n_e$ and $n_H$ are the electron and hydrogen densities, respectively. $dR$ and $dA$ are the radial and
area integral elements, respectively. Since $n_e \propto n_H$, then
$$
F_{UV} \propto \tau_{radio} \>dA. 
$$

The combination of radio and UV flux data, especially when 
spatially resolved, i.e., when $dA$ is measured, provides very powerful constraints on  
inhomogeneous atmospheric structures because of the very different temperature sensitivities of radio and
collisionally excited UV line emission.

\section{Astrophysical Impact}

One of the major goals of this research is to be able to build a quantitative predictive model of atmospheric
structure and mass loss based on underlying physical principles. Such a model could be used to improve stellar 
and galactic chemical evolution calculations, and predict the mass-loss histories of RSG which would in turn
help to interpret early-time supernova spectra \citep{moriya2017}.
Today we are very far from such a model.
The observational constraints that the ngVLA will provide on red giant and RSG extended atmospheres, 
where the winds accelerate, would be a major step forward. 
ngVLA would provide the basic measurements of size, electron density scale-height, 
and degree of uniformity that any theoretical model must satisfy.

\section{Uniqueness of the ngVLA Capabilities}

The power of the ngVLA lies in its high-spatial resolution {\em and} high sensitivity which, when combined, 
provide the ability to probe the spatial scales of the extended near-star atmospheres of evolved stars. 
This is the zone where the physics that controls different aspects of stellar activity will reveal itself.

\acknowledgements GMH thanks CU-CASA for infrastructure resources to support ongoing stellar astrophysics 
radio research. This research has made use of NASA's Astrophysics Data System Bibliographic 
Services and the VizieR catalogue access tool, CDS,
 Strasbourg, France \citep{vizier2000}.



\end{document}